\documentstyle[sprocl,epsfig]{article}

\def\Journal#1#2#3#4{{#1} {\bf #2}, #3 (#4)}

\def\NPA{{\em Nucl. Phys.} A}
\def\NPB{{\em Nucl. Phys.} B}
\def\PLB{{\em Phys. Lett.}  B}
\def\PRL{\em Phys. Rev. Lett.}
\def\PRD{{\em Phys. Rev.} D}
\def\PRC{{\em Phys. Rev.} C}
\def\ZPA{{\em Z. Phys.} A}
\def\ZPC{{\em Z. Phys.} C}

\def\be{\begin{equation}}
\def\ee{\end{equation}}
\def\bea{\begin{eqnarray}}
\def\eea{\end{eqnarray}}
\def\alt{\vcenter{\vbox{\hbox{$\buildrel < \over \sim$}}}}
\def\agt{\vcenter{\vbox{\hbox{$\buildrel > \over \sim$}}}}
\def\preprint#1{\vskip -180pt \noindent\hfill\hbox{#1}\vskip 175pt}

\begin{document}

\title{HARD EXCLUSIVE PHOTOPRODUCTION \\ OF $\Phi$ AND $J/\Psi$ MESONS
\footnote{Talk given by W. Schweiger at the joint INT/Jefferson Lab workshop on\lq\lq
Exclusive and Semi-Exclusive Processes at High Momentum Transfer\rq\rq, Newport News, USA,
May 1999}}

\author{C. F. BERGER}

\address{Department of Physics and Astronomy, SUNY Stony Brook\\ Stony Brook, 
NY 11794-3800, USA \\E-mail: carola.berger@sunysb.edu} 

\author{W. SCHWEIGER}

\address{Institut f\"ur Theoretische Physik, Universit\"at Graz, 
Universit\"atsplatz 5\\ A-8010 Graz, Austria \\E-mail:
wolfgang.schweiger@kfunigraz.ac.at}

\maketitle
\preprint{UNIGRAZ-UTP 07-09-99}
\abstracts{We consider the reaction $\gamma\, p\, \rightarrow \, M\, p$, with $M$ being
either a $\Phi$ or a  $J/\Psi$ meson, within perturbative QCD treating the proton as a 
quark-diquark system. The phenomenological couplings of gauge bosons to (spatially extended)
diquarks and the quark-diquark distribution amplitude of the proton are adopted from
previous investigations of baryon form factors and two-photon processes. Going beyond
leading order, we take into account hadron-mass effects by a systematic expansion in the
small parameter (hadron mass/ photon energy). With the meson distribution amplitudes taken
from the literature our predictions for the differential cross section at
$\vert t \vert\, \agt\, 3\, \hbox{GeV}^2$ seem to provide a reasonable extrapolation of the
low-$t$ data in case of the $\Phi$. They fail, however, completely for the $J/\Psi$. A closer
inspection reveals that \lq\lq hadron-mass corrections\rq\rq\  become dominant for $t$
values of only a few GeV in case of the $J/\Psi$. This indicates that one has to go to much
larger energies and momentum transfers in order to achieve a reliable description of
$J/\Psi$ production within the hard-scattering approach.}

\section{Introduction}
The forthcoming data from the 93-031 experiment at CEBAF will provide precise information
on exclusive photoproduction of vector mesons ($\rho, \omega, \Phi$) in the
momentum-transfer range $1\, \alt\, \vert t \vert\, \alt\, 5$~GeV$^2$. For this kinematics 
non-perturbative (vector-meson dominance) and perturbative (quark and gluon exchange)
production mechanisms are supposed to compete. From the pertur\-bation-theoretical viewpoint
the $\Phi$ channel is certainly the cleanest and simplest of these three production channels.
The valence Fock state of the $\Phi$, i.e. the $s$-$\bar{s}$ state, can only be produced via
two-gluon exchange. Quark exchange is OZI suppressed if the strangeness content of the
nucleon is small. Both quark and gluon exchanges are possible, however, in the case of the
$\rho$ and the $\omega$. In the present contribution we are interested in describing $\Phi$
production in the perturbative regime. We employ a modified version of the hard-scattering
approach (HSA)~\cite{BL89} in which the proton is considered as a quark-diquark system. This
kind of perturbative model has already been applied successfully to a number of exclusive
photon-induced reactions~\cite{Ja93}$^-\,$\cite{KSPS97}. A consistent description of baryon
electromagnetic form factors, Compton scattering off baryons, etc. has been achieved in the
sense that the corresponding large momentum-transfer data ($p_\perp^2\, \agt\, 3$~GeV$^2$)
are reproduced with the same set of model parameters. In addition to $\Phi$ production we
investigate whether the same model could be applied to $\gamma\, p\, \rightarrow \, J/\Psi\,
p$ in the few-GeV momentum-transfer region, where experimental measurements seem to be
feasible in the future. In the following we start with a short outline of the diquark model,
say a few words about helicity amplitudes for photoproduction of vector mesons and their
scaling behavior within our model. Finally, we compare our differential cross-section
results for photoproduction of
$\Phi$s and $J/\Psi$s with the existing (low-$t$) data and with predictions of other models
and draw our conclusions.

\section{The Hard-Scattering Approach with Diquarks}
\label{sec:diquark}
\begin{figure}[t!]
\begin{center}
\epsfig{file=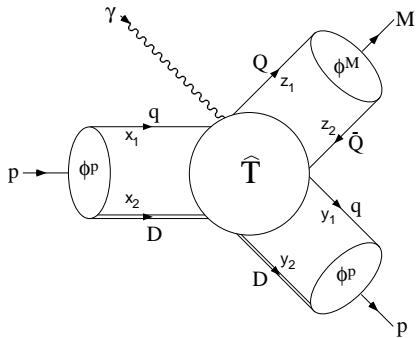,height=4.5cm,clip=}
\end{center}
\caption{Graphical representation of the  hard-scattering formula, Eq.~(\ref{convol}),
for $\gamma \, p\, \rightarrow\, M \, p$. $x_i$, $y_i$, and $z_i$ denote
longitudinal momentum fractions of the constituents.}
\label{kinem}
\end{figure}

\begin{figure}[t!]
\begin{center}
\epsfig{file=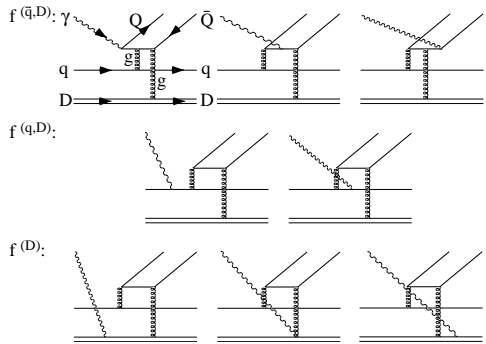,height=4.5cm,clip=}
\end{center}
\caption{Eight of the, altogether, sixteen tree-level diagrams which contribute to the
elementary process $\gamma\,q\, D\,  \rightarrow\, Q \, \bar{Q}\, q \, D$. The
remaining diagrams are obtained by interchanging the two gluons. We have also indicated how
the diagrams are grouped to gauge invariant expressions.}
\label{feyn}
\end{figure}

Within the hard-scattering approach a helicity amplitude $M_{\{\lambda\}}$ for
the reaction $\gamma \, p\, \rightarrow\, M \, p$ is (to
leading order in $1/p_\perp$) given by the convolution integral~\cite{BL89}
\begin{equation}
M_{\{\lambda\}}(\hat{s},\hat{t}) \! = \! \int_0^1 \! \! dx_1 dy_1 dz_1
{\phi^M}^{\dagger}(z_1,\tilde{p}_{\! \perp})
{\phi^p}^{\dagger}(y_1,\tilde{p}_{\! \perp})
\widehat{T}_{\{\lambda\}}(x_1,y_1,z_1;\hat{s},\hat{t})
\phi^p(x_1,\tilde{p}_{\! \perp}) \, . \label{convol}
\end{equation}
The distribution amplitudes (DAs) $\phi^H$ are probability amplitudes for
finding the valence Fock state in the hadron $H$ with the constituents carrying certain
fractions of the momentum of their parent hadron and being  collinear up to a
factorization scale $\tilde{p}_{\perp}$. In our model the valence Fock state of an ordinary
baryon is assumed to consist of a quark ($q$) and a diquark ($D$). We fix our notation in
such a way that the momentum fraction appearing in the argument of $\phi^{H}$ is
carried by the quark -- with the momentum fraction of the other constituent (either
diquark or antiquark) it sums up to 1 (cf. Fig.~\ref{kinem}). For our actual
calculations the (logarithmic) $\tilde{p}_{\perp}$ dependence of the DAs is neglected
throughout since it is of minor importance in the restricted energy and
momentum-transfer range we are  interested in. The hard scattering amplitude
$\widehat{T}_{\{\lambda\}}$ is calculated perturbatively in collinear approximation and
consists in our particular case of all possible tree diagrams contributing to the elementary
scattering process  $\gamma\,q\, D\,  \rightarrow\, Q \, \bar{Q}\, q \, D$. Eight of the,
altogether, sixteen diagrams which contribute to photoproduction of (heavy)
quarkonia are depicted in Fig.~\ref{feyn}. The subscript ${\{\lambda\}}$ of $\widehat{T}$
represents a set of possible photon, proton and vector-meson helicities. The Mandelstam
variables $s$ and $t$ are written with a hat to indicate that they are defined for vanishing
hadron masses. 
Our calculation of the hard-scattering amplitude involves an expansion in
powers of $(m_H/\sqrt{\hat{s}})$ which is performed at fixed scattering angle. We keep only
the leading order and next-to-leading order terms in this expansion. Hadron masses, however, are
fully taken into account in flux and phase-space factors.

The model, as applied in Refs.~2-5,
comprises scalar ($S$) as well as axial-vector ($V$) diquarks.
$V$-diquarks become relevant if one wants to describe spin observables
which require the flip of baryonic helicities. For the Feynman rules
of electromagnetically and strongly interacting diquarks, as well as
for the choice of the quark-diquark DAs of octet
baryons we refer to Ref.~2.
Here it is only important to mention that the composite nature of diquarks is taken into
account by multiplying each of the Feynman diagrams entering the hard-scattering amplitude
with diquark form factors. These are parameterized by  multipole functions  with the power
chosen in such a way that in the limit $p_{\perp} \rightarrow \infty$ the scaling behavior
of the pure quark HSA is recovered. We restrict the (running) strong coupling constant
$\alpha_s$ to be smaller than $0.5$. For further details of the diquark model we refer to 
the publication of R.~Jakob et al.~\cite{Ja93}. The numerical values of the model parameters
for the present study are also taken from this paper.

The only new ingredient which appears in the photoproduction of mesons is the meson DA.
A model for the $\Phi$-meson DA which fulfills QCD sum-rule constraints has been
proposed by Benayoun and Chernyak~\cite{BC90}. In order to treat $\Phi$ and $J/\Psi$
production on the same footing we adapt this DA by attaching an exponential factor which
provides a flavor dependence. The effect of this factor is negligible for $\Phi$s, but it
becomes crucial for $J/\Psi$s.  The form of the exponential factor is inspired by a simple
relativistic treatment of heavy mesons~\cite{BSW85} and provides a flavor dependence in
accordance with heavy-quark effective theory. For longitudinally polarized
$\Phi$s and $J/\Psi$s the DA reads
\begin{equation}
\phi^M (z_1,\lambda_\Phi=0) \propto z_1 (1 - z_1) \exp \left[ \tilde{b}^2 m_{M}^2 (z_1 -
0.5)^2
\right]\, .
\label{DABC}
\end{equation}
The DA for transversially polarized $\Phi$s and $J/\Psi$s contains an additional polynomial
in $z_1$ which enhances the maximum at $z_1=0.5$ and makes it narrower (cf.
Ref.~6). For the oscillator parameter $\tilde{b}$ we take a value of $0.97$ in
accordance with estimates for the radius of the $J/\Psi$ meson (see, e.g., the discussion in
Ref.~8). The absolute \lq\lq normalization\rq\rq\ of the DA, Eq.~(\ref{DABC}), is
related to the experimentally determinable decay constant of the $\Phi$ and $J/\Psi$ meson,
respectively.

At the end of this section we want to make two technical remarks. The first remark
concerns the numerical treatment of the convolution integral, Eq.~(\ref{convol}), which is
not straightforward since propagator singularities are encountered in the range of
integration. We have carefully separated the propagator singularities and integrated
analytically so that the remaining integrals could be performed by a rather fast fixed-point
Gaussian quadrature. For details of this integration procedure we refer to Ref.~5.
The second remark concerns the treatment of hadron masses. These are taken into account in
the hard-scattering amplitudes by assigning to every constituent of the hadron $H$ an effective mass
$x m_H$, where $x$ is the fraction of the four-momentum of the hadron $H$ carried by the
constituent. Keeping in mind that the momentum fractions are weighted by the hadron DA in
the convolution integral, Eq.~(\ref{convol}), this means that, e.g., the quarks  acquire (on
the average) a mass which is rather the mass of a constituent quark than a current quark.
This may be interpreted in such a way that at intemediate momentum transfers a quark
surrounded by a cloud of $q\bar{q}$ pairs and gluons is not resolved but rather acts like a
single particle with a corresponding mass. For a more detailed explanation of our treatment
of mass-effects which, by the way, also guarantees the gauge invariance of the
mass-correction terms we refer to Ref.~9.

\section{Helicity Amplitudes}
For exclusive photoproduction of vector mesons $\gamma\, p\, \rightarrow\, M\, p$ one finds,
altogether, 24 complex helicity amplitudes. By virtue of parity invariance only 12 of these
helicity amplitudes are independent. Following the notation of Ref.~10 we denote
them by
\begin{eqnarray}
H_{1,\lambda_\phi}&=&M_{\lambda_\phi,\lambda_f=+1/2,\lambda_\gamma=1,\lambda_i=-1/2} \, , 
\quad
H_{2,\lambda_\phi} = M_{\lambda_\phi,\lambda_f=+1/2,\lambda_\gamma=1,\lambda_i=+1/2} \, , 
\nonumber \\
H_{3,\lambda_\phi}&=&M_{\lambda_\phi,\lambda_f=-1/2,\lambda_\gamma=1,\lambda_i=-1/2} \, , 
\quad
H_{4,\lambda_\phi} = M_{\lambda_\phi,\lambda_f=-1/2,\lambda_\gamma=1,\lambda_i=+1/2} \, ,
\nonumber \\
\end{eqnarray}
with $\lambda_\phi = 0, \pm1$. For our normalization of the helicity amplitudes the 
unpolarized differential cross section takes on the form
\begin{equation}
\frac{d\sigma}{dt} = \frac{1}{32 \pi (s - m^2_{p})^2}
\sum_{\lambda_\phi=0,\pm 1} \sum_{i=1}^4 \vert H_{i,\lambda_\phi} \vert^2
\, .
\label{unpolwq}
\end{equation}

Within the HSA the energy dependence of the helicity amplitudes at 
fixed cm angle and large $s$ is roughly
\begin{eqnarray}
H_{2,0}\, &,& \,  H_{3,0} \, \propto s^{-5/2} \, , \nonumber \\
H_{1,0}\, &,& \,  H_{4,0} \, , \, H_{2,1}\, , \,  H_{3,1}  , \, H_{2,-1}\, , \,  H_{3,-1} \, \propto
s^{-3} \, ,
\nonumber \\ 
H_{1,1}\, &,& \,  H_{4,1}  , \, H_{1,-1}\, , \,  H_{4,-1} \, \propto s^{-7/2} \, , 
\label{scaling}
\end{eqnarray}
depending on whether the helicity of the hadronic constituents is conserved or flipped by
one or two units. This scaling behavior is modified by  logarithms due to the running
coupling $\alpha_s$ and, eventually, the evolution of the hadron distribution amplitudes.
Further  deviations are due to the diquark form factors. For small momentum tranfers
diquarks appear nearly point-like and the decay behavior of the helicity amplitudes is
weakened. The energy dependence of Eqs.~(\ref{scaling}) is only approached at large enough
momentum transfer, where the diquark form factors become fully operational.

If one neglects mass effects altogether, the helicity-flip amplitudes vanish and only
the hadron-helicity conserving amplitudes $H_{2,0}$ and $H_{3,0}$ are non-zero. Due to
(anti)symmetry properties of the Feynman graphs under interchange of the momentum fractions
$z_1$ and $z_2$, only diagrams in which the photon couples to the heavy quark contribute in
the massless case. Mass corrections suppressed by the order $(m_H/\sqrt{\hat{s}})$ do not
contribute to the non-flip amplitudes, but provide for non-vanishing single-flip
amplitudes. In order to get a non-vanishing result for the double-flip amplitudes it would
even be necessary to consider mass corrections suppressed by order $(m_H/\sqrt{\hat{s}})^2$.
Since such corrections would also show up in the helicity-conserving amplitudes and we do
not want to make things too complicated, we neglect them completely.

\section{Results and Conclusions}
\begin{figure}[ht!]
\begin{center}
\epsfig{file=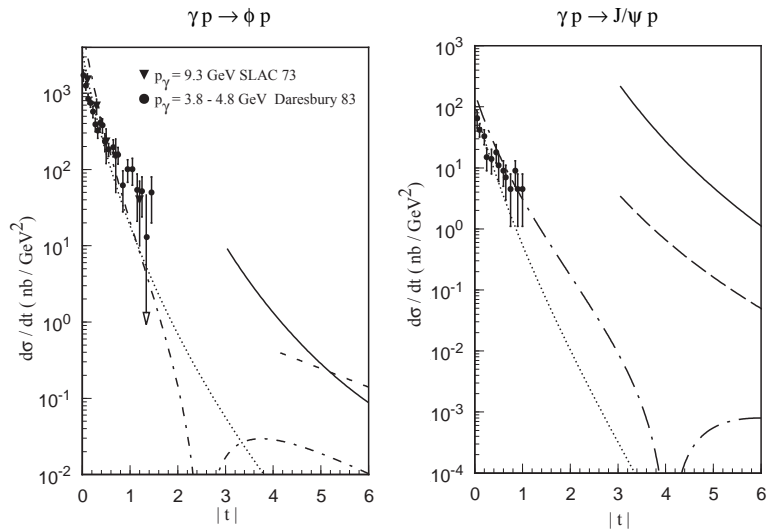,height=7.0cm,clip=}
\end{center}
\caption{Differential cross section for the photoproduction of the
$\phi$-$p$ ({\em left figure}) and the $J/\Psi$-$p$ ({\em right figure}) final
states for photon lab-energies of 6 and 150 GeV, respectively. The solid line corresponds to 
the diquark-model prediction. Cross sections resulting from the Pomeron-exchange
mechanism~{\protect \cite{DoLa87}} (dotted line) and a two-gluon-exchange model~{\protect
\cite{La95}} (dash-dotted line) are plotted for comparison. The short-dashed line in the left
figure represents an attempt to include correlations into the two-gluon-exchange
model~{\protect \cite{La98}}. The long-dashed line in the right figure has been obtained within
the diquark model by neglecting mass corrections, i.e. by taking into account only the
hadron-helicity conserving amplitudes $H_{2,0}$ and $H_{3,0}$.
$\phi$-data are taken from Ballam et al.~{\protect \cite{Bal73}} and Barber et al.~{\protect
\cite{Ba83}}. The $J/\Psi $ data points have been obtained by Binkley et
al.~{\protect \cite{Bi82}} by averaging the measured cross sections over photon
lab-energies between 60 and 300 GeV.}
\label{result}
\end{figure}

Fig.~\ref{result} shows (unpolarized) differential cross sections for the photoproduction
of $\Phi$ (left) and $J/\Psi$ (right) mesons. In the absence of large-$t$ data we are not able
to directly confront the diquark-model predictions (solid line) with experiment. However, comparing
with the experimental points which are only available up to $t\, \alt\, 1.5$~GeV$^2$, one can recognize
that our results provide a reasonable extrapolation in case of the
$\Phi$, but seem to fail completely for the $J/\Psi$. This failure can be traced back to the
\lq\lq hadron-mass corrections\rq\rq\ which enter the cross section only via the
hadron-helicity-flip amplitudes (apart from flux and phase-space factors). Whereas the two
hadron-helicity conserving amplitudes provide already about 60\%  of the cross section in
case of the $\Phi$, they are negligible in case of the $J/\Psi$ (solid line vs. long-dashed
line). This means for the $J/\Psi$ that hadron-mass effects become dominant for the
kinematics considered in the plot. Thus they are not reliably estimated by keeping only the
lowest order terms in a mass expansion. By comparing the kinematics of the two plots in
Fig.~\ref{result} one observes that the relevant parameter of our mass expansion, namely 
$(m_M/\sqrt{\hat{s}})$, is in both cases small. Both reactions are considered within the same
momentum-transfer range, but within very different angular ranges. $\vert t \vert$ values
between 3 and 6~GeV$^2$ correspond to a scattering angle between 70 and 110 degrees for a
photon lab-energy of 6~GeV, but to nearly-forward scattering (between 10 and 20 degrees) for
a photon lab-energy of 150~GeV. Keeping in mind that our mass expansion is an
expansion in the parameter $(m_H/\sqrt{\hat{s}})$ at {\bf fixed} angle, the different
magnitude of mass corrections in $\Phi$ and $J/\Psi$ production becomes clearer. The
expansion coefficient of the first mass-correction term is angular dependent and large for
small scattering angles. This indicates that perturbation theory for
$J/\Psi$ production is only trustworthy at large enough energies and scattering angles.

The problem that $\Phi$ production is well reproduced, whereas $J/\Psi$ production is
overestimated does not only occur within our model for the hard-scattering region, but is
also encountered if one describes the diffractive scattering by means of a
pomeron-exchange mechanism~\cite{DoLa87} (in which the pomeron directly couples to the
quarks) or the two-gluon-exchange model proposed by Laget and Mendez-Galain~\cite{La95}. The
latter model represents an attempt to describe the transition from diffractive to hard
scattering. Within this model the two gluons may be considered as a remnant of the pomeron.
For both models the situation in $J/\Psi$ production is improved by arguing that the coupling
of the pomeron to a charmed quark is weaker than the coupling to a strange quark.
Predictions of these models are also depicted in Fig~\ref{result}. The two-gluon-exchange
model exhibits a characteristic node which, however, is completely washed out if
correlations between the protonic constituents are taken into account, i.e. if the two gluons
are allowed to couple to different constituents of the proton. For $\Phi$ production and
$\vert t \vert\, \agt\, 4$~GeV$^2$ our results and the results of the two-gluon-exchange
model {\bf with} correlations become comparable. This is not surprising, since correlations
are automatically contained in our HSA-based model. 

To conclude, within our perturbative diquark model we are not able to make reliable
predictions for exclusive photoproduction of $J/\Psi$ mesons in the few-GeV
momentum-transfer range where experimental data can be expected. The reason is that
effects due to the charm-quark mass are only reliably estimated (within our approach) at
very large energies and momentum tranfers. For exclusive photoproduction of $\Phi$ mesons,
however, our results provide a reasonable extrapolation of the low-$t$ data and are also
comparable with a two-gluon exchange model which continues the pomeron exchange for
diffractive scattering to higher momentum transfers. CEBAF, or even better, an upgrade of
CEBAF, could help to figure out how different mechanisms (pomeron exhange $\rightarrow$
two-gluon exchange $\rightarrow$ hard-scattering mechanism) in high-energy photoproduction
evolve when going from low to high $t$. Our model will meet the first severe test in the
near future when the data analysis of the CEBAF 93-031 experiment for $\gamma\, p\, 
\rightarrow \, \Phi\, p$ will be completed.

\end{document}